\documentclass[a4paper,12pt]{article}
\usepackage[english]{babel}

\def\beq{\begin{equation}}
\def\eeq{\end{equation}}
\def\bea{\begin{eqnarray}}
\def\eea{\end{eqnarray}}
\def\zbar{\bar{z}}
\def\Jbar{\bar{J}}
\def\bra{\langle}
\def\ket{\rangle}
\def\sa{\left[}
\def\sk{\right]}
\def\pat{\partial}

\begin{document}

\pagestyle{plain}

\vspace{1cm}
\begin{titlepage}
\begin{flushleft}
       \hfill                      {\tt hep-th/0110252}\\
       \hfill                      HIP-2001-56/TH \\
       \hfill                      October 27, 2001\\
\end{flushleft}
\vspace*{3mm}
\baselineskip18pt
\begin{center}
{\Large {\bf The Spectrum of Strings on BTZ Black Holes \\
and Spectral Flow in the SL(2,R) WZW Model }} \\
\vspace*{12mm}
{\large
Samuli Hemming\footnote{E-mail: samuli.hemming@hip.fi} \\
Esko Keski-Vakkuri\footnote{E-mail: esko.keski-vakkuri@hip.fi}\\}

\vspace{5mm}

{\em Helsinki Institute of Physics \\
P.O. Box 9\\
FIN-00014  University of Helsinki \\
Finland }
\vspace*{10mm}
\end{center}


\begin{abstract}
We study the spectrum of bosonic string theory on rotating BTZ
black holes, using a SL(2,R) WZW model.
Previously, Natsuume and Satoh have analyzed strings
on BTZ black holes using orbifold techniques. We show how an appropriate
spectral flow in the WZW model
can be used to generated the twisted sectors, emphasizing how the
spectral flow works in the hyperbolic basis natural for the BTZ
black hole. We discuss the projection condition which leads to the
quantization condition for the allowed quantum numbers for the string
excitations, and its connection to the anomaly in the
corresponding conserved Noether current.
\end{abstract}

\end{titlepage}

\baselineskip16pt

\section{Introduction}

One of the obstacles in testing the AdS/CFT correspondence \cite{M,agmoo} in its
strong form has been gaps in understanding string theory in AdS
backgrounds. In the case of AdS$_3$, the spacetime can be
identified as the $SL(2,R)$ group manifold, and string theory can
be formulated as a $SL(2,R)$ WZW model \cite{bunch}. It was recently understood
\cite{MO} that the model has a spectral flow symmetry which can be used to
understand better the string spectrum, and to study the consistency of the
model \cite{MOS}.

One interesting extension is to study Lorentzian orbifolds of the AdS$_3$
spacetime. Conical defects in AdS$_3$ spacetimes, or
point masses in AdS$_3$ \cite{DJH}, for a discrete set of deficit angles or
masses can be constructed as AdS$_3/Z_N$ orbifolds; and black
holes in AdS$_3$ (the Ba\~{n}ados-Teitelboim-Zanelli (BTZ) black
holes \cite{BTZ}) can be constructed as AdS$_3/Z$ orbifolds.
Apart from being interesting in their own right, they appear in
near-horizon geometries of (spinning) six-dimensional black strings,
five-dimensional black holes, or extremal four-dimensional black holes
\cite{nearhor}. More precisely, the near-horizon
geometries are (fibered) BTZ$\times S^3$ geometries. Since strings
on $S^3$ can be studied with $SU(2)$ WZW models, it might be
possible to formulate string theory in the above near-horizon geometries
as a combination of $SL(2,R)$ and $SU(2)$ WZW models, in the spirit of
\cite{GPS}. Moreover, it
has been emphasized recently that conical defects in AdS$_3$ and
BTZ black holes give a simple setting where to study formation of
black holes in string theory \cite{conestring}.

Strings on AdS$_3/Z_N$ orbifolds were recently investigated in the context of
spectral flow \cite{Son,MM}.
Among other things, it was shown how the twisted sectors can be thought to
be generated by spectral flow. String theory on BTZ black holes
was investigated by Natsuume and Satoh \cite{NS}, using a WZW model
based on the
$\widetilde{SL}(2,R)/Z$ orbifold interpretation \cite{HW, Kaloper}.
They constructed the twisted sector
vertex operators, and analyzed the properties of the spectrum. In
the light of current understanding \cite{MO}, ref. \cite{NS} focused on the
properties of
short strings. At that time the understanding of the string
spectrum was incomplete. There appeared to
be ghosts in the spectrum, unless the spectrum was truncated by hand.
But such a truncation implied an upper bound on the mass of the
string states. These puzzles were clarified
in \cite{MO}, where it was found that the spectrum must also
include long strings, generated by spectral flow from tachyonic
untwisted strings.

In this paper, we reinterpret the string spectrum on rotating BTZ black holes,
in particular the twisted sector, as being generated by spectral
flow. We show explicitly how the spectral flow is realized, and
discuss the projection needed for the level matching condition.
Following \cite{GPS,NS}, we interpret the problem of finding the
projection generator as a Noether current ambiguity, and discuss
that in some more detail than in \cite{NS}. We do not address the
problem of ghosts, or modular invariance. We believe it will be
straightforward to generalize the corresponding results from the
work in pure AdS$_3$ to the present set-up. Here the focus has
been in setting up the dictionary.

\section{BTZ Black Hole and the WZW model}

We begin from the WZW action:
\beq
S = \frac{k}{8\pi} \int d^2 \sigma {\rm Tr} \left( g^{-1} \partial_a
g g^{-1} \partial^a g \right) + \frac {ik}{12\pi} \int {\rm Tr}
\left( g^{-1} dg \wedge  g^{-1} dg \wedge  g^{-1} dg \right) \ .
\eeq
An element $g$ of the group $SL(2,R)$ can be parametrized as
\bea
g &=& \frac 1{\ell^2} \left( \begin{array}{cc}
x_1 + x_2 & x_3 + x_0 \\
x_3 - x_0 & x_1 - x_2
\end{array} \right) \\
\ell^2 &=& x_0^2 + x_1^2 - x_2^2 - x_3^2
\eea
The latter equation shows that the manifold $SL(2,R)$ is a
three-dimensional hyperboloid embedded in $R^{2,2}$. In fact,
this is exactly same for $AdS_3$, so they are the same manifold.
The dimensional parameter $\ell$ is then the $AdS_3$
radius\footnote{We will henceforth work in units where $\ell=1$.}.
In contrast to the $SU(2)$ WZW model, the level $k$ is not required
to be an integer since the cohomology class $H^3$ vanishes
for $SL(2,R)$.

We introduce a new coordinate system in $AdS_3$ in order to obtain
the BTZ black hole. The $AdS_3$ manifold is divided into three
regions \cite{BTZ}, that descibe the regions outside the
outer horizon, between the outer and the inner horizon, and inside
the inner horizon. In this paper we will focus on strings in
the exterior region of the black hole, and leave the issue of including all the
regions across the horizon(s) for further study. The exterior
region, outside the outer horizon
($\hat r^2 > 1$), is described by the following
coordinate patch:
\bea
x_1 &=& \hat r \ {\rm cosh} \, \hat \phi \\
x_2 &=& \hat r \ {\rm sinh} \, \hat \phi \\
x_0 &=& \sqrt{\hat r^2 - 1 } \ \ {\rm sinh} \, \hat t \\
x_3 &=& \sqrt{\hat r^2 - 1 } \ \ {\rm cosh} \, \hat t
\eea
The corresponding group elements can be expressed using the
generators of $SL(2,R)$ in the hyperbolic basis \cite{NS}:

\beq
g = e^{u\sigma_3}\, e^{\rho\sigma_1}\, e^{-v\sigma_3}
\label{hypbas}
\eeq
where we have introduced new coordinates $u$, $v$ and $\rho$:
\bea
\hat r &=&  {\rm cosh} \, \rho \\
\hat t &=& u + v \\
\hat \phi &=& u -v
\eea

With this parametrization the action looks like
\beq
S = \frac 1{4\pi \alpha '} \int d^2\sigma \sqrt{h} \left(
h^{\alpha\beta} G_{\mu\nu} + i \epsilon^{\alpha\beta}B_{\mu\nu} \right)
\partial_{\alpha} X^{\mu} \partial_{\beta} X^{\nu} \, ,
\label{action_btz} \eeq which is the Polyakov action for a string
propagating in a spacetime in the presence of a background field
$B_{\mu\nu}$. The spacetime metric of this action reads \beq ds^2
= \alpha' k\left\{ -\left( \hat r^2 - 1 \right) d\hat t^2 + \frac
{d\hat r^2}{\hat r^2 - 1} + \hat r^2 d\hat
\phi ^2 \label{metric_hats} \right\} \eeq and the antisymmetric background
field is \beq B = \alpha' k \hat r^2 \, d\hat \phi \wedge
d\hat t \ . \eeq
A further change of variables \bea \hat r^2 &=&
 \left( \frac{r^2 - r^2_-}{r^2_+ - r^2_-} \right) \\
\left(
\begin{array}{c} \hat t \\ \hat \phi \end{array} \right) &=&
 \left( \begin{array}{cc} r_+ & -r_- \\ -r_- & r_+
\end{array} \right) \left( \begin{array}{c} t \\ \phi \end{array}
\right)
\eea
brings the metric into a form where the black hole is easily recognized:
\beq
ds^2 = \alpha' k \left\{ - \frac{(r^2 - r^2_+)(r^2-r^2_-)}{r^2}dt^2
+  \frac{r^2}{(r^2 - r^2_+)(r^2-r^2_-)}dr^2
+ r^2(d\phi - \frac{r_+r_-}{r^2}dt)^2 \right\} \ .
\label{metric_nohats}
\eeq
The antisymmeric tensor is
\beq
 B = \alpha' k r^2 d\phi \wedge dt \ ,
\eeq
up to an exact two form. After the periodic identification of the coordinate
\beq
\phi \sim \phi + 2 \pi \ ,
\label{per_id}
\eeq
one obtains the BTZ black hole. In the coordinates
(\ref{metric_hats}),
the periodicity condition reads as
\beq
(\hat t, \hat \phi ) \sim (\hat t - 2\pi r_- , \hat
\phi + 2\pi r_+ ) \ .
\eeq
The allowed values for the coordinates (\ref{metric_nohats})
are $-\infty <t<+\infty$,
$0\leq r <+\infty$, and $0\leq \phi < 2\pi$. By allowing the time
coordinate to have all real values, we have moved to the universal
cover $\widetilde{SL}(2,R)$ of $SL(2,R)$.
The positive constants
$r_{\pm}$ denote the radii of the outer and the inner horizon.
The mass $M_{BH}$ and the angular momentum $J_{BH}$
of the black hole can be read off from the metric (\ref{metric_nohats}):
\bea
M_{BH} &=& (r_+^2 + r_-^2) \\
J_{BH} &=& 2  r_+ r_-
\eea

\section{Geodesic equations of the BTZ black hole}

The geodesic equations for the BTZ metric have been derived
in \cite{CMP}:
\bea
r^2 \dot r^2 &=& -m^2 \left( r^4 - r^2 M_{BH}
+ \frac{J_{BH}}{4} \right) + \left( E^2 - L^2
\right) r^2 + L^2 M_{BH} - E L J_{BH} \\
\dot \phi &=& \frac{ \left( r^2 - M_{BH} \right) L
+ \frac 12 E J_{BH}}
{r^4 - r^2 M_{BH} + \frac 14 J^2_{BH} } \\
\dot t &=& \frac{E r^2 - \frac 12 L J_{BH}}
{r^4 - r^2 M_{BH} + \frac 14 J^2_{BH} }
\eea
The constants $E$ and $L$ are associated with the Killing vectors of
the BTZ metric, $\partial_t$ and $\partial_\phi$.

For $m^2>0~(=0,<0)$, the solutions describe motions of timelike
(lightlike, spacelike)
particles. As explained in \cite{MO},
spectral flow will stretch the timelike geodesics to
worldsheets of short strings, and spacelike
geodesics to worldsheets of long strings. Incidentally, it should
be noted that the spacetime of a rotating black hole is
geodesically complete.

The stress tensor remains constant on geodesics.
The left- and rightmoving components of the stress tensor can be written in the form
\beq
  T_{L,R} = -2\alpha'~{\rm
  Tr}~(J_{L,R}J_{L,R}) \ ,
\eeq
where $\alpha'$ is the string slope.
Substituting the currents (see eqn. (\ref{curr})) and the group
elements (\ref{hypbas}), we obtain
\beq
  T_{R,L} = \frac{\alpha' k^2}{4} \left\{ (\pat_\alpha
  \hat{\phi})^2 \cosh^2 \rho  - (\pat_\alpha \hat{t})^2 \sinh^2 \rho
   + (\pat_\alpha \rho )^2 \right\}
\eeq
where $\pat_\alpha =\pat_+~(\pat_-)$ for $T_R~(T_L)$.
Then, for example, substituting a radial geodesic in a
non-rotating black hole,
\bea
 \dot{\hat{r}}^2 &=& -m^2 (\hat{r}^2-1)+\hat{E}^2 \nonumber \\
 \dot{\hat{t}} &=& \frac{\hat{E}}{\hat{r}^2-1}
\eea
we obtain
\beq
   T_L = T_R = -\frac{\alpha' k^2 m^2}{4} \equiv \mp
   \frac{1}{4}k\alpha^2 \ ,
\eeq
which is of the same form as in \cite{MO}, after identifying $\alpha^2 =
|km^2|$. The components $T_{R,L}$ are negative (positive) for timelike
(spacelike) geodesics.

\section{Strings on BTZ}

The WZW model has a chiral $\widetilde{SL}(2,R)_R\times \widetilde{SL}(2,R)_L$
symmetry, which implies the existence of conserved currents:
\beq
J_R (x^+) = \frac {ik}2 \, \partial_+ \, g \, g^{-1} , \ \
J_L (x^-) = \frac {ik}2 \, g^{-1} \, \partial_- \, g
\label{curr}
\eeq
The generators of $SL(2,R)$ are $\tau^0 = - \frac 12 \sigma_2$,
$\tau^1 = \frac i2 \sigma_1$ and
$\tau^2 = \frac i2 \sigma_3$.
With the hyperbolic parametrization (\ref{hypbas}), the
components
\beq
    J^a_{L,R} = -2~{\rm Tr}~(\tau^a J_{L,R})
\eeq
read as
\bea
J^2_R &=& k \left( \partial_+ u - {\rm cosh}\, 2\rho \,
\partial_+ v \right) \label{curr_param_r} \\
J^{\pm}_R &=& k \left( \pm \partial_+ \rho - {\rm sinh}\, 2\rho \,
\partial_+ v \right) e^{\mp 2u} \\
J^2_L &=& k \left( -\partial_- v + {\rm cosh}\, 2\rho \,
\partial_- u \right) \label{curr_param_l} \\
J^{\pm}_L &=& k \left( \pm \partial_- \rho - {\rm sinh}\, 2\rho \,
\partial_- u \right) e^{\mp 2v}
\eea
where $J^{\pm}$ are defined as $J^{\pm} = J^0 \pm J^1$. This is
contrast to the elliptic parametrization, where the $J^{\pm}$ are defined as
linear combinations of the non-compact generators $J^1,J^2$.

The Hilbert space of the WZW model decomposes into products of
irreducible unitary representations of the $\widehat{SL}_k(2,R)$ current algebras.
Using the mode expansions
\beq
  J^a_R = \sum^{\infty}_{n=-\infty} \ J^a_n \ e^{-inx^+}  \ ; \
  J^a_L = \sum^{\infty}_{n=-\infty} \ \bar{J}^a_n \ e^{-inx^-} \ ,
\eeq
the non-trivial current commutation relations become
\bea
   \left[ J^2_n, J^{\pm}_m \right] &=& \pm i J^{\pm}_{n+m} \\
   \left[J^+_n, J^-_m \right] &=& -2iJ^2_{n+m} -kn\delta_{n+m,0}
   \label{jpmcomm} \\
   \left[J^2_n,J^2_m \right] &=& \frac{k}{2} n\delta_{n+m,0}
\eea
and similarly for the antiholomorphic sector. The currents and the
Virasoro generators are found to satisfy the commutation relations
\bea
   \left[ L_n, J^a_m \right] &=& -mJ^a_{n+m} \\
   \left[ L_n, L_m \right] &=& (n-m)L_{n+m} +\frac{c}{12}
   n(n^2-1)\delta_{n+m,0} \ .
\eea

The currents and the group elements satisfy the OPE's
\beq
   J^a(z)g(w,\bar{w}) \sim \frac{-\tau^a g}{z-w} \ ; \
   \bar{J}^a (\zbar ) g(w,\bar{w}) \sim \frac{-g\tau^a}{\zbar -\bar{w}}
\eeq
which yields the commutation relations
\beq
  \left[ J^a_n, g\right] = -\tau^a g w^n \ ; \
  \left[ \Jbar^a_n , g\right] = -g\tau^a \bar{w}^n \ .
\label{jgcomm}
\eeq
Using the parametrization (\ref{hypbas}) and the relations (\ref{jgcomm}), one
can derive the time translation and space rotation generators, satisfying
\bea
  \delta_t g &=& i\delta t \left[ Q_t , g\right] \\
  \delta_\phi g &=& i\delta \phi \left[ Q_{\phi} ,g\right] \ ,
\eea
to be the following combinations of zero modes\footnote{Note however
that the definition of the generators has a Noether ambiguity, to be
discussed in section 5.} of $J^2,\Jbar^2$:
\bea
  Q_t  &=& \Delta_-J^2_0 -\Delta_+ \Jbar^2_0 \\
  Q_\phi &=& \Delta_- J^2_0 +\Delta_+ \Jbar^2_0 \ .
\eea

In \cite{NS}, Natsuume and Satoh discussed the spectrum of strings
on BTZ black holes. They first constructed the Kac-Moody
primaries transforming under the unitary irreducible
representations of the global $\widetilde{SL}(2,R)_R\times \widetilde{SL}(2,R)_L$
symmetries,
using the hyperbolic basis. The vertex operators for
the primaries are of the form
\beq
   V^{j,0}_{J_R,J_L} = D^j_{J_R,J_L} (g) e^{-iJ_Ru +iJ_L v}
\eeq
where $D^j_{J_R,J_L}(g)$ are matrix elements of the unitary irreps
of the $\widetilde{SL}(2,R)$ group. The representations that
appear in the Hilbert space are the principal discrete
representations (short strings)
and the principal continuous representations (long strings). A subtlety, that we
will discuss at more length below, is that the representations are
now expressed in the hyperbolic basis which diagonalizes the
non-compact generators $J^2_0,\bar{J}^2_0$.

To describe the
black hole, one needs to incorporate the periodicity of the
angular coordinate $\phi$.
So, one needs to twist the WZW model with respect to the discrete
action of $Q_\phi$ which generates the periodic
identifications. After constructing the twist fields $W_n(z,\zbar)$ \cite{DFMS} which
create twisting with winding number $n$, one obtains the
generic twisted sector primary fields
\beq
    V^{j,n}_{J_R,J_L}(z,\zbar) = V^{j,0}_{J_R,J_L}(z,\zbar )
    W_n(z,\zbar) \ .
\eeq
The Kac-Moody primaries are then the states
\beq
   | j,J_R,n\ket | j, J_L,n \ket = V^{j,n}_{J_R,J_L}|0\ket|0\ket \
   .
\eeq

\subsection{Representations in the hyperbolic basis}

The unitary irreducible representations of $SL(2,R)$ are typically
discussed in the elliptic basis which diagonalizes the non-compact
direction. When the commutation relations are written in the
hyperbolic basis,
\beq
  \sa J^+_0,J^-_0 \sk = -2iJ^2_0 \ ; \
  \sa J^2_0 , J^{\pm}_0 \sk = \pm i J^{\pm}_0
\label{comrel}
\eeq
they appear first slightly puzzling, because the latter equation
seems to contradict the hermiticity of $J^2_0$. Let us introduce
\cite{KMS} a basis of eigenvectors of $J^2_0$:
\beq
   J^2_0 |\lambda \ket = \lambda |\lambda \ket \ ; \
   \bra \lambda | \lambda' \ket = \delta (\lambda -\lambda') \ ,
\eeq
where $\lambda \in R$ since $J^2_0$ is Hermitean.
Then, it appears that the state $J^+_0 | \lambda \ket$
is an eigenstate of $J^2$ with eigenvalue $\lambda + i$. The
solution to the puzzle is the same as for the Heiseberg
algebra $[p,q]=-i$ in quantum mechanics: the basis vectors $|\lambda \ket$
do not represent normalizable vectors in the Hilbert space.
A normalizable vector is a linear combination
\beq
    | \phi \ket = \int^{\infty}_{-\infty} d\lambda \ \phi
    (\lambda) |\lambda \ket
\eeq
where the wavefunction $\phi (\lambda)$ must satisfy
\beq
  \| \phi \| \equiv \int^{\infty}_{-\infty} d\lambda |\phi
  (\lambda)|^2 < \infty \ .
\eeq
Each of the generators $J_0^2,J^{\pm}_0$ has its corresponding
domain of vectors $|\phi \ket$ on which it is defined. For
example, for $J^+_0$ the wavefunction $\phi (\lambda)$ determines
via analytic continuation a unique new wavefunction
\beq
    J^+_0|\phi\ket = J^+_0 \int^{\infty}_{-\infty} d\lambda \ \phi
    (\lambda) |\lambda \ket = \int^{\infty}_{-\infty} d\lambda \ f(\lambda )
    \phi (\lambda -i) |\lambda \ket
\label{jphi}
\eeq
such that one obtains a normalizable state:
\beq
   \int^{\infty}_{-\infty} d\lambda \ |f(\lambda )\phi (\lambda
   -i)|^2 < \infty \ .
\eeq
One could formally introduce states labeled as $|\lambda +i\ket$
to rewrite the relation (\ref{jphi}). However, these states should be
understood to be expanded in the original basis $\{|\lambda
\ket\}$ where the $J^2_0$ eigenvalues are real.
Similarly, the operator $J^-_0$ acts as
\beq
 J^-_0|\phi\ket = J^-_0 \int^{\infty}_{-\infty} d\lambda \ \phi
    (\lambda) |\lambda \ket = \int^{\infty}_{-\infty} d\lambda \ g(\lambda +i)
    \phi (\lambda +i) |\lambda \ket \ .
\eeq
The functions $f(\lambda ),g(\lambda)$ play the role of matrix
elements of $J^+_0,J^-_0$. The commutation relation (\ref{comrel}) requires
that
\beq
     f(\lambda)g(\lambda)-g(\lambda +i)f(\lambda -i) = -2i\lambda
     \ .
\eeq
Now, to construct the irreducible representations of $SL(2,R)$
using the hyperbolic basis, one views them as reducible
representations with respect to the subalgebra generated by
$J^2_0,J^+_0$. Irreducible representations of the latter algebra are given
in the basis $|\lambda\ket$. So one needs to find out how these
representations are added together to form the irreducible representations of
$SL(2,R)$. In particular, one needs to identify which one of the
known irreducible representations is being constructed. This is
done by checking that the spectrum of
the eigenvalues of the compact generator $J^0_0=(J^+_0+J^-_0)/2$
corresponds to that of the continuous or discrete series
representations.
Note that both of $J^{\pm}_0$ are hermitian operators, instead
of conjugates of one another, so it turns out that one of the
matrix element functions, {\em e.g.} $f(\lambda )$, can be chosen freely,
with the other one $g(\lambda )$ being determined by the commutation
relation (\ref{jpmcomm}).
Particular choices of $f(\lambda)$ then turn out to correspond to a
discrete or continuous spectrum of $J^0_0$, and yield the discrete
or continuous representations of $SL(2,R)$ \cite{KMS}. This will
also determine the multiplicity of the representations of the
subalgebra.

So the upshot is that in the hyperbolic basis, the states of the
irreducible representations of $SL(2,R)$ are of the form
\beq
   |\lambda, r\ket \ ; \ J^2_0 |\lambda ,r\ket = \lambda |\lambda
   ,r \ket
\eeq
where $r$ is an index that enumerates the multiplicities of the
representations of the $J^2_0,J^+_0$ subalgebra.

\subsection{Representations of the current algebra}

After having obtained the representations of $\widehat{SL}(2,R)$, or the
Kac-Moody primaries, the rest of the states in the current
algebra (the representation of $\widehat{SL}_k(2,R)$) are obtained
by acting on the states with the generators $J^{2,\pm}_{-n}$, $n\geq
1$:
\beq
     J_N |J_R ,r \ket \bar{J}_N |J_L ,r \ket
\eeq
where $J_N,\bar{J}_N$ denote generic products of $J^a_n$'s and
$\bar{J}^a_n$'s and we are back to using the labels
$J_R,J_L$ for the continuous real eigenvalues
of $J^2_0,\Jbar^2_0$. However, to keep the $J^2_0$ eigenvalues real,
one must restrict the number of $J^+_{-n}$'s in $J_N$ to be equal
to the number of $J^-_n$'s \cite{NS}. One then considers only states of the
form
\beq
     K_N |J_R ,r \ket \bar{K}_N |J_L, r\ket
\eeq
where $K_N$ is a generic product of operators $K^a_{-n}$ defined by
\beq
   K^2_{-n} \equiv J^2_{-n} \ ; \ K^+_{-n} \equiv J^+_{-n}J^-_0
   \ ; \ K^-_{-n} \equiv J^-_{-n}J^+_0 \ .
\eeq
They satisfy the commutation rules
\beq
   \sa J^2_0,K^a_{-n} \sk = 0 \ ; \ \sa L_0 , K^{\pm}_{-n} \sk =
   nK^{\pm}_{-n} \ .
\label{kcomm}
\eeq

\subsection{Spectral flow}

Next we consider the spectral flow \cite{MO}. We take a solution of
the equations of motion for the WZW model and generate a new one
using a specific coordinate transformation.
In the hyperbolic basis, the spectral flow for $SL(2,R)$ model
reads:
\beq
g \rightarrow e^{-i w_+ x^+ \tau^2}\, g\,
e^{i w_- x^- \tau^2}
\eeq
so the coordinates $u$, $v$ transform as $u \rightarrow u + \frac 12
w_+ x^+$, $v \rightarrow v + \frac 12 w_- x^-$. In particular, the
time and angular coordinates transform as
\bea
 \hat{t} &\rightarrow& \hat{t} + \frac 12 [(w_+ +w_-)\tau + (w_+ -w_-)\sigma
 ] \\
 \hat{\phi} &\rightarrow& \hat{\phi} + \frac 12 [(w_+ +w_-)\sigma + (w_+
 -w_-)\tau ] \ .
\eea
So, after the periodic identifications (\ref{per_id}) which make a BTZ black
hole, the spectral flow parameters $w_{\pm}$ are allowed to have
the discrete values
\beq
      w_{\pm} = (r_+ \mp r_-)n \equiv \Delta_{\mp} n \ .
\label{wpm}
\eeq

Under spectral flow, the components of the currents transform:
\bea
J^2_R &\rightarrow& \tilde J^2_R \equiv J^2_R + \frac k2 w_+ \ ; \
J^{\pm}_R \ \rightarrow \  \tilde J^{\pm}_R \equiv J^{\pm}_R e^{\mp w_+ x^+} \\
J^2_L &\rightarrow& \tilde J^2_L\equiv J^2_L - \frac k2 w_- \ ; \
J^{\pm}_L \ \rightarrow \ \tilde J^{\pm}_L\equiv J^{\pm}_L e^{\mp w_- x^-}
\eea
and the stress tensor transforms as
\bea
 T_R&\rightarrow& \tilde T_R = T_R + w_+ J^2_R + \frac k4 (w_+)^2 \\
 T_L&\rightarrow& \tilde T_L = T_L - w_- J^2_L + \frac k4 (w_-)^2 \ .
\eea

After spectral flow in the hyperbolic basis, the $J^2$
component of the current is still periodic:
\beq J^2 (x^+2\pi) = J^2(x^+)
\eeq
leading to the expansion
\beq
J^2 (z) = \sum_{s \in Z} z^{-s-1} J^2_s \ .
\eeq
The $J^{\pm}$ components become quasiperiodic:
\beq
J^{\pm} (x^+ + 2\pi) = e^{\mp 2\pi w_+} J^{\pm} (x^+)
\eeq
giving rise to the expansions
\beq
J^{\pm} (z) = \sum_{s \in Z \pm iw_+} z^{-s-1} J^{\pm}_s \ .
\eeq
The antiholomorphic current satisfies the same properties and
equations, with $w_+$ replaced by $w_-$.
The nonvanishing ommutator relations of the flowed current generators are the same
as in (37)-(39), with the mode labels $n,m$ replaced by
complex labels $r,s\in Z\pm iw_+$.

{}From above,  we see that under spectral flow
the components of $J^2$, $\bar J^2$ transform as:
\bea
J^2_n &\rightarrow& \tilde J^2_n \equiv J^2_n + \frac k2 \, w_+ \delta_{n,0}
\ ; \ J^\pm_n \ \rightarrow \ \tilde J^\pm_n \equiv J^\pm_{n\pm iw_+} \\
\bar J^2_n &\rightarrow& \tilde{\bar{J}}^2_n \equiv
\bar J^2_n - \frac k2 \, w_- \delta_{n,0}
\ ; \ \bar J^\pm_n \ \rightarrow \ \tilde{\bar{J}}^\pm_n \equiv
\bar J^\pm_{n\pm iw_-} \ \ .
\eea
We also find the following transformation rules for the Virasoro
generators:
\bea
L_n &\rightarrow& L_n + w_+ J^2_n + \frac k4 w_+^2 \delta_{n,0} \\
\bar L_n &\rightarrow& \bar L_n - w_- \bar J^2_n + \frac k4 w_-^2
\delta_{n,0}
\eea
The commutation relations of the current algebra and
the Virasoro algebra remain invariant under spectral flow.
Further, the commutation relations (\ref{kcomm}) also remain invariant under
spectral flow. The generators $K^a_n$ transform under spectral
flow as
\beq
    K^2_n \rightarrow \tilde K^2_n = \tilde J^2_n \ ; \
    K^\pm_n \rightarrow \tilde K^\pm_n = \tilde J^\pm_n \tilde
    J^\mp_0 = J^\pm_{n\pm iw_+} J^\mp_{n\mp iw_+} \ ,
\eeq
and it is easy to check that
\beq
   \sa \tilde J^2_0,\tilde K^a_{-n} \sk = 0 \ ; \
   \sa \tilde L_0 , \tilde K^{\pm}_{-n} \sk =
   n\tilde K^{\pm}_{-n} \ .
\eeq

The Virasoro constraints and the spectrum are similar to those in \cite{Son},
except for the allowed values of $w_{\pm}$. The $L_0$ constraint
for the holomorphic part is
\beq
\left( L_0 - 1 \right) \left| \tilde j , \tilde{J}_R \right> =
\left( -\frac {\tilde j (\tilde j + 1)}{k-2} + h + \tilde N
+ w_+ \tilde{J}_R - \frac k4 w_+^2 - 1 \right) \left| \tilde j ,
\tilde{J}_R \right> = 0
\eeq
and for the antiholomorphic part:
\beq
\left( \bar L_0 - 1 \right) \left| \tilde j , \tilde{J}_l \right> =
\left( -\frac {\tilde j (\tilde j + 1)}{k-2} + \bar h + \tilde {\bar N}
- w_- \tilde{J}_L - \frac k4 w_-^2 - 1 \right) \left| \tilde j ,
\tilde{J}_L \right> = 0 \ .
\eeq
These lead to the following allowed values for $\tilde{J}_R$,
$\tilde{J}_L$:
\bea
\tilde{J}_R &=& \frac k4 w_+ - \frac 1{w_+} \left( - \frac {
\tilde j (\tilde j - 1)}{k-2} + h + \tilde N -1 \right) \\
\tilde{J}_L &=& - \frac k4 w_- + \frac 1{w_-} \left( - \frac
{\tilde j (\tilde j - 1)}{k-2} + \bar h + \tilde {\bar N} -1 \right)
\eea
The result agrees (up to conventions) with \cite{NS}.
The level matching condition requires that
\beq
   \tilde{\bar{N}}_{tot} -\tilde{N}_{tot} = w_+
   \tilde{J}_R
    + w_- \tilde{J}_L - \frac{k}{4} (w^2_+ -w^2_-)
    = {\rm integer} \ ,
\eeq
where $\tilde{\bar{N}}_{tot},\tilde{N}_{tot}$ are the total level
numbers. Substituting the allowed discrete values (\ref{wpm}) for the
spectral flow parameters, leads to the conditions
\bea
     \tilde{N}_{tot} &=& \tilde{\bar{N}}_{tot} + nl \\
       l &=& \Delta_- \tilde{J}_R + \Delta_+ \tilde{J}_L
     +\frac{k}{2}nJ_{BH} \label{levelm}
\eea
where $n$ is the winding number and $l\in Z$ is the angular
momentum. With the latter condition, the vertex operators can be
shown to be periodic \cite{NS}.

Let us finally comment on the space-time energy spectrum of the
string. The short strings correspond to the principal discrete
series representations. They are parameterized by the spin $j$
which can take any real value $j>0$. The eigenvalue of the compact
generator $J^0_0$ is of the form $j+q$, where $q$ is an integer.
For strings in AdS$_3$ vacuum, one uses the basis where $J^0_0$ is
diagonalized, and it is possible to solve for $j$ as a function of
the level of the current algebra, the level number of string
excitations, and the spectral flow parameter $w$ which takes
discrete values. So one finds that only a discrete set of spin
values $j$ are realized in the string spectrum. In other words, a
discrete set of principal discrete representations are realized in
the spectrum of short string excitations. Thus, the spacetime
energy spectrum of short strings is discrete.

However, in the BTZ black hole background, the representations are
expressed in the hyperbolic basis. The short strings again
correspond to the principal discrete series representation, with
discrete $J^0_0$ eigenvalues $j+q$. However, now it is the non-compact
operator $J^2_0$ which has been diagonalized. The equations (84)
and (85) relate the eigenvalues of the non-compact operators to
the level $k$, the level numbers $N,\bar{N}$ and the spin $j$.
But these are not related to the $J^0_0$ eigenvalues. So there is
no equation to solve for the spin $j$ as a function of just the discrete
parameters $k,N,w$. Hence we conclude that all principal discrete
representations $j>0$ are realized in the string spectrum. Then,
the $J^2_0,\bar{J}^2_0$ eigenvalues can take any real value
$J_L,J_R$. The two continuous parameters are related by one
constraint equation (\ref{levelm}), leaving one free continuous
parameter. The spacetime energies of short string excitations are
related to the eigenvalues of the time translation
generator $Q_t$ given by (46). So the energy
spectrum
$$
   E= \Delta_-J_R - \Delta_+ J_L
$$
is continuous, as opposed to the discrete spectrum in the AdS$_3$
case. But a continuous spectrum is just what we expected to find in the
black hole case.

\section{Invariant subspace}

The physical spectrum consists of states that are invariant under
the periodic identification $\phi \sim \phi + 2 \pi$. This imposes
a condition on the allowed quantum numbers. The projection is
realized by the operator
\beq
      P = \exp \left( i2\pi Q \right) \ ,
\label{P}
\eeq
where $Q$ is the operator that generates translations in
$\phi$. The states that are invariant under the action by $P$ are to be
retained in the spectrum. This requires the eigenvalues $l$ of
$Q$ to be integers
on the states that are retained. The eigenvalue $l$ corresponds to
angular momentum.

However, there is a subtlety in finding out the correct generator
$Q$. There is a well known ambiguity in defining the Noether
current in field theory, since one is always free to add to it
a divergence of an antisymmetric tensor without affecting the
conservation law. However, this may give an additional
contribution to the conserved charge in a topologically
non-trivial sector. This was discussed in the context of WZW coset
models {\em e.g.} in \cite{GPS, NS}. The conserved current
must be such that the charge that appears in the projection
operator yields a projection condition consistent with
the level matching condition.

Naively, one might consider the operator $Q_\phi = \Delta_-J^2_0+\Delta_+\Jbar^2_0$
to be the generator to be used in the projection operator. This
would yield the quantization condition
\beq
    \Delta_- \tilde{J}_R + \Delta_+ \tilde{J}_L + knJ_{BH} = l
    \in Z \
\eeq
on the states which are kept under the projection. However, if the
black hole is rotating, $J_{BH}\neq 0$, this is clearly inconsistent with the
level matching condition (\ref{levelm}) -- there is a mismatch by a factor of
2 in the last term, for strings with non-zero winding number $n$.

Another way to derive the conserved current under rotations in
$\phi$ is to use the sigma model action (\ref{action_btz}).
Varying with respect to $\phi$ gives rise to the conserved current
with the time component
\beq
  J_{\tau} = \frac{k}{2\pi} \{r^2 \pat_\tau \phi - \frac{1}{2}
  J_{BH}\pat_\tau t -(r^2-c) \pat_\sigma t \} \ .
\label{J_1}
\eeq
Here $c$ is an arbitrary constant, reflecting the freedom to add
an exact two-form to $B$.

Let us compare this with the current $\Delta_-J^2_R+\Delta_+J^2_L$
using the expressions (\ref{curr_param_r}), (\ref{curr_param_l}):
\bea
 J_{\phi} &\equiv & \Delta_-J^2_R+\Delta_+ J^2_L \nonumber \\
\mbox{} &=& k\{ r^2~\pat_\tau \phi
  -\frac{1}{2}J_{BH} \pat_\tau t - (r^2-M_{BH}) \pat_\sigma t
  - \frac{1}{2}J_{BH} \pat_\sigma \phi \} \ . \nonumber
\eea
This agrees with (\ref{J_1}), up to the last total derivative term.
The last term plays a role in the winding sector, where $\pat_\sigma \phi
=n\neq 0$. In the non-winding sector the currents agree,
and give rise to the same symmetry generator
\beq
    Q_{action} = \int^{2\pi}_0 d\sigma~J_\tau = Q_\phi = \Delta_-J^2_0 +
    \Delta_+\Jbar^2_0 = \int^{2\pi}_0 \frac{d\sigma}{2\pi}~J_\phi
    \ .
\eeq
However, in the topologically non-trivial winding sector,
the current ambiguity plays a role, and contributes to the current
generator. The correct generator which agrees
with the level matching condition turns out to be $Q_{action}$.
From above, we see that generator $Q$ to use in $P$ is
\bea
  Q = Q_{action} &=& \int^{2\pi}_0 \frac{d\sigma}{2\pi}~(J_\phi
   +\frac{k}{2}J_{BH}\pat_\sigma \phi ) \nonumber \\
  \mbox{} &=& \Delta_- \tilde{J}^2_0 + \Delta_+ \tilde{\Jbar}^2_0
  + \frac{k}{2}nJ_{BH} \ ,
\eea
leading to the quantization condition
\beq
  \Delta_- \tilde{J}_R + \Delta_+ \tilde{J}_L +
  \frac{k}{2}nJ_{BH}= l \in Z \ ,
\eeq
in agreement with (\ref{levelm}).

\bigskip

\noindent
{\bf \large Acknowledgments and Note Added}

\bigskip

E.K-V. has been in part supported by the Academy of Finland, and
thanks Hirosi Ooguri for a useful discussion. S.H. has been in
part supported by the Magnus Ehrnrooth foundation.
As we were finalizing the paper, there appeared another paper
\cite{FH} which also examines spectral flow and BTZ black holes,
but there the spectral flow appears in a different role.

\end{document}